\documentclass[11pt]{article}
\usepackage{url,fullpage,graphicx,amssymb,amsmath}

\title{Fast $(1+\epsilon)$-approximation of the L\"owner extremal matrices of  high-dimensional symmetric matrices}

\author{Frank Nielsen\footnote{Frank Nielsen is with \'Ecole Polytechnique
and Sony Computer Science Laboratories Inc. {\tt Frank.Nielsen@acm.org}
{\tt Frank.Nielsen@acm.org} } 
\and
Richard Nock\footnote{Richard Nock is with NICTA \& ANU, Australia.}
}

\date{February 2011\\
(online \today)}

\def\inner#1#2{ \langle {#1},{#2} \rangle }
\def\innerproduct#1#2{ \langle {#1},{#2} \rangle }

\def\trace{\mathrm{tr}}
\def\tr{\mathrm{tr}}

\def\vec{\mathrm{vec}}

\newtheorem{problem}{Problem}

\newtheorem{theorem}{Theorem}

\def\calL{\mathcal{L}}
\def\calA{\mathcal{A}}
\def\calB{\mathcal{B}}
\def\calS{\mathcal{S}}
\def\calC{\mathcal{C}}
\def\calF{\mathcal{F}}
\def\calO{\mathcal{O}}

\def\calE{\mathcal{E}}

\def\Sym{\mathrm{Sym}}
\def\diag{\mathrm{diag}}

\def\ball{\mathrm{Ball}}

\def\ext{\mathrm{ext}}

\def\bbR{\mathbb{R}}
\def\bbC{\mathbb{C}}
\def\st{\ : \ }

\def\vech{\mathrm{vech}}
\def\CH{\mathrm{CH}}

\def\Int{\mathrm{Int}}

\begin{document}
\maketitle 
 
\begin{abstract}
Matrix data sets are common nowadays like in biomedical imaging where the Diffusion Tensor Magnetic Resonance Imaging (DT-MRI) modality produces data sets of
 3D symmetric positive definite matrices anchored at voxel positions capturing the anisotropic diffusion properties of water molecules in biological tissues.
The space of symmetric matrices can be partially ordered using the L\"owner ordering, and computing extremal matrices dominating a given set of matrices is a basic primitive used in matrix-valued signal processing. 
In this letter, we design a fast and easy-to-implement iterative algorithm to approximate arbitrarily finely these extremal matrices.
Finally, we discuss on extensions to matrix clustering.
\end{abstract}

{\bf keywords} :
Positive semi-definite matrices, L\"owner ordering cone, extremal matrices, geometric covering problems, core-sets, clustering.

\section{Introduction: L\"owner extremal matrices  and their applications}
Let $M_d(\bbR)$ denote the space of {\em square} $d\times d$ matrices  with real-valued coefficients, 
 and $\Sym_d(\bbR)=\{ S \st S=S^\top\}\subset M_d(\bbR)$ the matrix vector space\footnote{Although addition preserves the symmetric property, beware that the product of two symmetric matrices may be not symmetric.} of {\em symmetric}  matrices.
A matrix $P\in M_d(\bbR)$ is said {\em Symmetric Positive Definite}~\cite{Bathia-2009} (SPD, denoted by $P\succ 0$) iff. $\forall x\not =0, x^\top P x >0$ and only
 {\em Symmetric Positive Semi-Definite}\footnote{Those definitions extend to Hermitian matrices $M_d(\bbC)$.} 
(SPSD,  denoted by $P\succeq 0$) when we relax the strict inequality ($\forall x, x^\top P x \geq 0$).
Let $\Sym^+_d(\bbR)=\{X \st X\succeq 0\}\subset \Sym_d(\bbR)$ denote the space of positive semi-definite matrices,
and $\Sym^{++}_d(\bbR)=\{X \st X\succ 0\}\subset \Sym^+_d(\bbR)$ denote the space of positive definite matrices.
A  matrix $S\in\Sym_d(\bbR)$ is defined by $D=\frac{d(d+1)}{2}$ real coefficients, and so is a SPD or a SPSD matrix.
Although $\Sym_d(\bbR)$ is a {\em vector space}, the SPSD matrix space does not have the vector space structure but is rather an abstract {\em pointed convex cone} with {\em apex} the zero matrix $0\in\Sym^+_d(\bbR)$ since  $\forall P_1,P_2\in \Sym^+_d(\bbR), \forall\lambda\geq 0, \quad P_1+\lambda P_2\in \Sym^+_d(\bbR)$.
Symmetric matrices can be {\em partially} ordered using the {\em L\"owner   ordering}:\footnote{Also often written Loewner in the literature, e.g., see~\cite{AppLoewner-1967}.}
  $P\succeq Q  \Leftrightarrow  P-Q\succeq 0$ and $P\succ   Q  \Leftrightarrow  P-Q\succ  0$.
When $P\succeq Q$,  matrix $P$ is said to {\em dominate} matrix $Q$, or equivalently that matrix $Q$ is dominated by matrix $P$. 
Note that the difference of two SPSD matrices may not be a SPSD matrix.\footnote{For example, consider $P=\diag(1,2)$ and $Q=\diag(2,1)$ then $P-Q=\diag(-1,1)$ and $Q-P=\diag(1,-1)$.}
A non-SPSD symmetric matrix $S$ can be dominated by a SPSD matrix $P$ when $P-S\succ 0$.\footnote{
For example, $S=\diag(-1,1)$ is dominated by $P=\diag(1=|-1|,1)$ (by taking the absolute values of the eigenvalues of $S$).}

The {\em supremum} operator  is defined on $n$ symmetric matrices $S_1, \ldots, S_n$   (not necessarily SPSDs) as follows:
\begin{problem}[L\"owner maximal matrices]
\begin{equation}
\bar S= \inf\{ X\in\Sym(\bbR)\ :\ \forall i\in [n], X\succeq S_i \},
\end{equation}
where $[n]=\{1, ...,n\}$.
\end{problem}
This matrix $\bar S=\max(S_1,\ldots, S_n)$ is indeed the ``smallest'', meaning the {\em tightest upper bound}, since by definition there does not exist another symmetric matrix $X'$ dominating all the $S_i$'s and dominated by $\bar S$.
Trivially, when there exists a matrix $S_j$ that dominates all others of a set $S_1, \ldots, S_n$, then the supremum of that set is matrix $S_j$.
Similarly, we define the {\em minimal/infimum matrix} $\underline{S}$ as the tightest lower bound.
 Since matrix inversion reverses the L\"owner ordering ($A\succ B \Leftrightarrow B^{-1} \succ A^{-1}$),
we link those extremal supremum/infimum matrices when considering sets of invertible symmetric matrices as follows:
$\underline{S}= \left (\max(S_1^{-1}, ..., S_n^{-1}) \right)^{-1}$.
Extremal matrices are {\em rotational invariant} $\max (O^\top S_1 O, \ldots , O^\top S_n O) = O^\top \times \max(S_1, \ldots , S_n) \times O$, where $O$ is any orthogonal matrix ($OO^\top=O^\top O=I$). This property is important in DT-MRI processing that should be invariant to the chosen reference frame.

Computing  L\"owner extremal matrices are useful in many applications: For example, in matrix-valued imaging~\cite{Angulo-2013,PSDMorpho-2007} (morphological operations, filtering, denoising or image pyramid representations), in formal software verification~\cite{QI-switchedsystems-2015}, in statistical inference with domain constraints~\cite{LownerMLE-1991,MLE-Lowner-2007}, in structure tensor of computer vision~\cite{Forsner-1986} (F{\"o}rstner-like operators), etc. 

This letter is organized as follows: Section~\ref{sec:covering} explains how to transform the extremal matrix problem into an equivalent geometric minimum enclosing ball of balls. Section~\ref{sec:mebb} presents a fast iterative approximation algorithm that scales well in high-dimensions. Section~\ref{sec:concl} concludes by hinting at further perspectives.

\section{Equivalent geometric covering problems\label{sec:covering}}

We build on top of~\cite{MatrixData-2007} to prove that solving the $d$-dimensional L\"owner maximal matrix amounts to either find (1) the minimal covering L\"owner matrix cone (wrt. set containment $\subseteq$) of a corresponding sets of $D$-dimensional cones (with $D=\frac{d(d+1)}{2}$), or (2) the minimal enclosing ball of a set of corresponding $(D-1)$-dimensional ``matrix balls''  that we cast into a geometric {\em vector} ball covering problem for amenable computations.
 
\subsection{Minimal matrix/vector cone covering problems}

Let $\calL=\{X\in\Sym^+(d) \st X\succeq 0\}$ denote the {\em L\"owner ordering cone} , and $\calL(S_i)$ the reverted and translated {\em dominance cone} (termed the penumbra cone in~\cite{MatrixData-2007}) with apex $S_i$ embedded in the space of symmetric matrices that represents all the symmetric matrices dominated by $S_i$: 
$\calL(S_i)  = \{ S\in\Sym_d(\bbR) \st S_i\succeq S\} =  S_i \ominus \calL$, where $\ominus$ denotes the {\em Minkowski set subtraction operator}: 
$\calA\ominus \calB=\{ a-b \st a\in \calA, b\in \calB\}$ (hence, $\calL(0)=-\calL$).
A matrix $S$ dominates $S_1, \ldots, S_n$ iff. $\forall i\in [n], \calL(S_i)\subseteq \calL(S)$.
In plain words, $S$ dominates a set of matrices iff. its associated dominance cone $\calL(S)$ covers all the dominance cones $\calL(S_i)$ for $i\in [n]$.
The dominance cones are ``abstract'' cones  defined in the $d\times d$ symmetric matrix space that can be ``visualized'' as equivalent {\em vector} cones in dimension $D=\frac{d(d+1)}{2}$ using  {\em half-vectorization}:
For a  symmetric matrix $S$, we stack the elements of the lower-triangular matrix part of $S=[s_{i,j}]$ (with $s_{i,j}=s_{j,i}$):
$\vech(S)=\left[s_{1,1}\ \ldots s_{d,1} \ s_{2,2}\ \ldots\ s_{d,2} \ \ldots\ s_{d,d} \right]^\top\in \bbR^(\frac{d(d+1)}{2})$.
Note that this is not the unique way to half-vectorize symmetric matrices but it is enough for {\em geometric containment} purposes.
Later, we shall enforce that the $\ell_2$-norm of  vectors $\vech(S)$ matches the Fr\"obenius matrix norm $\|\cdot\|_F$.

Let $\calL_v$ denotes the vectorized matrix L\"owner ordering cone: $\calL_v=\{\vech(P) \st P\succ 0\}$, and $\calL_v(S)$ denote the {\em vector dominance cone}: 
$\calL_v(S)=\{\vech(X) \st X\in \calL(S)\}$. 
Next, we further transform this minimum $D$-dimensional matrix/vector cone covering problems as   equivalent {\em Minimum Enclosing Ball} (MEB) problems of $(D-1)$-dimensional matrix/vector balls.

\subsection{Minimum enclosing ball of ball problems}

A {\em basis} $\calB$ of a convex cone $\calC$ anchored at the origin $0$ is a convex subset $\calB\subseteq \calC$ so that $\forall x\not =0\in \calC$ 
there exists a {\em unique decomposition}: $x=\lambda b$ with $b\in \calB$ and $\lambda>0$.
For example, $\Sym^+_1(\bbR)=\{ P\in \Sym^+(\bbR) \st \tr(P)=1\}$ is a basis of the L\"owner cone $\calL=\Sym^+(\bbR)$.
Informally speaking, a  basis of a cone can be interpreted as a {\em compact} cross-section of the cone.
The L\"owner cone $\calL$ is a smooth convex cone   with its interior $\Int(\calL)$ denoting the space of positive definite matrices $\Sym^{++}(\bbR)$ (full rank matrices), and its border $\partial \calL=\calL\backslash \Int(\calL)$ the {\em rank-deficient}
symmetric positive semi-definite matrices (with apex the zero matrix $0$ of rank $0$).
A point $x$ is an {\em extreme element} of a convex set $S$ iff. $S\backslash\{x\}$ remains convex.
It follows from Minkowski theorem that every compact convex set $\calS$ in a finite-dimensional vector space can be reconstructed
as convex combinations of its extreme points $\ext(\calS)\subseteq\partial \calS$: 
That is, the compact convex set is the closed convex hull of its extreme points.

\begin{figure}
\centering
\includegraphics[width=0.88\columnwidth]{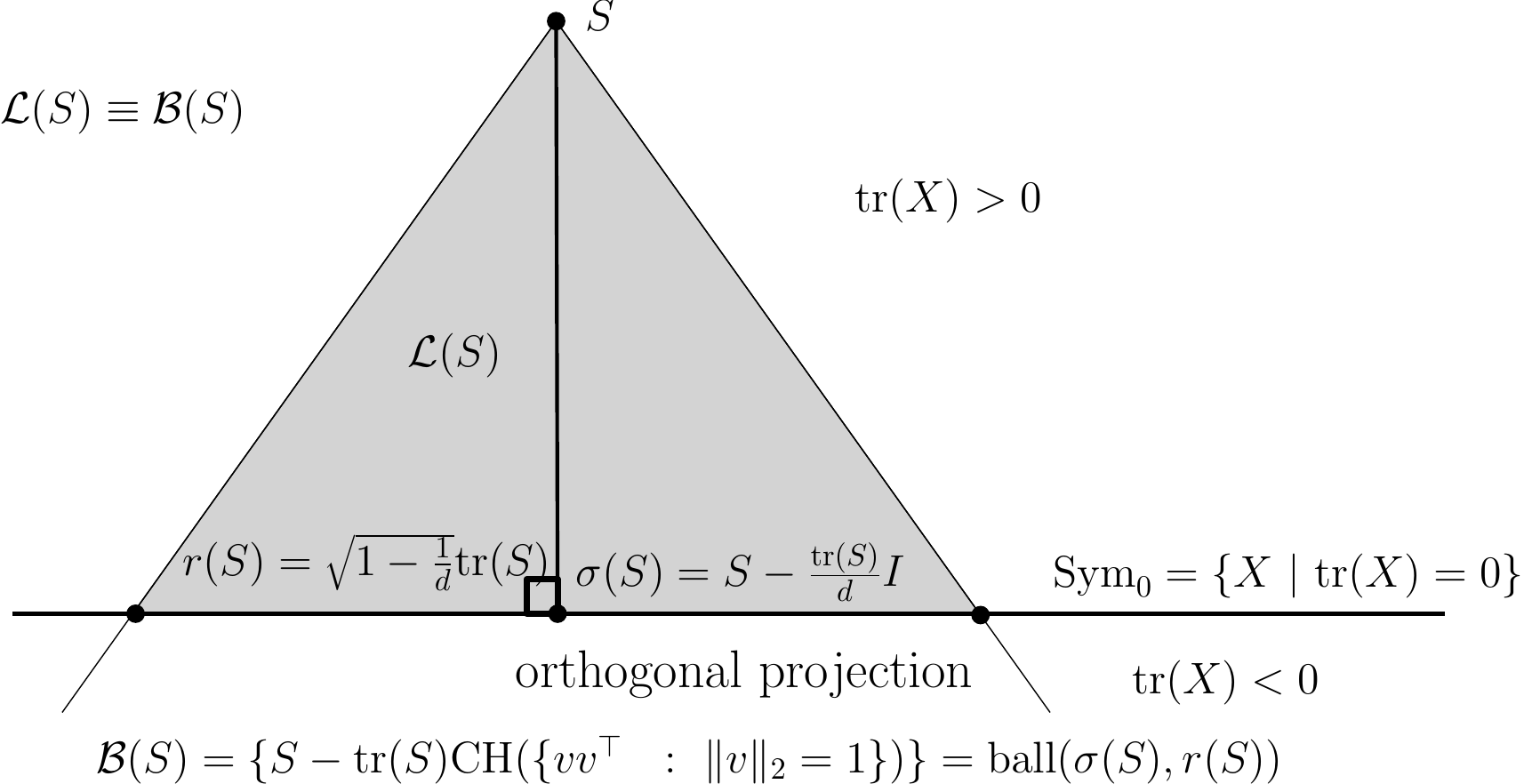}

\caption{The dominance cone $\calL(S)$ associated with matrix $S$ has apex $S$ and base $\calB(S)=\ball(\sigma(S),r(S))$, a ball centered at matrix $\sigma(S)$ of radius $r(S)$.
The cone $\calL(S)$ has an equivalent representation $\calB(S)$ provided that $\tr(S)\geq 0$.
\label{fig:coneprojection}}
\end{figure}

A face $\calF\subset \calC$ of a closed cone $\calC$ is a subcone such that $x+y \in \calF \rightarrow x,y \in \calF$.
The $1$-dimensional faces are the {\em extremal rays} of the cone.
The basis of the L\"owner ordering cone is~\cite{SPSDCone-1987} $\calB(\calC)=\CH(vv^\top \st v\in\bbR^d, \|v\|_2=1)$.
Other rank-deficient or full rank matrices can be constructed by convex combinations of these rank-$1$ matrices, the extremal rays.

For any square matrix $X=[x_{i,j}]$,  the {\em trace operator} is defined by $\tr(X)=\sum_{i=1}^d x_{i,i}$, the sum of the diagonal elements of the matrix.
The trace also amounts to the sum of the eigenvalues $\lambda_i(X)$ of matrix $X$: $\tr(X)=\sum_{i=1}^d \lambda_i(X)$.
The basis $\calB_i$ of a dominance cone $\calL(S_i)$ is $\calB_i=\{S_i-\tr(S_i)\times \calB(\calL)\}$.
Note that all the basis of the dominance cones lie in the {\em subspace} $H_0$ of symmetric matrices with zero trace.
Let $\inner{X}{Y}_F=\tr(X^\top Y)$ denote the {\em matrix inner product} and $\|M\|_F=\sqrt{\inner{M}{M}_F}=\sqrt{\sum_{i,j} m_{i,j}^2}$ the matrix {\em Fr\"obenius norm}. Two matrices $X$ and $Y$ are orthogonal (or perpendicular) iff. $\inner{X}{Y}_F=0$.
It can be checked that the identity matrix $I$ is perpendicular to any zero-trace matrix $X$ since $\innerproduct{X}{I}_F=\tr(X)=0$.
The center of the  ball basis of the dominance cone  $\calL=\calL(S)$ is obtained as the {\em orthogonal projection} of $S$ onto the zero-trace subspace $H_0$:
$\sigma(S) = S - \frac{\tr(S)}{d}I$.
The dominance cone basis is a {\em matrix ball} since for any rank-$1$ matrix $E=v v^\top$ with $\|v\|_2=1$ (an extreme point), we have the radius:

\begin{equation}\label{eq:matrad}
r(S) = \| S -\tr(S) vv^\top - \sigma(S) \|_F = \trace(S)\sqrt{1-\frac{1}{d}},
\end{equation}
that is non-negative since we assumed that $\tr(S)\geq 0$.
Reciprocally, to a basis ball $B=\ball(\sigma,r)$, we can associate the apex of its corresponding dominance cone 
$\calL(B)$: $\sigma + \frac{r}{d}\frac{I}{\sqrt{1-\frac{1}{d}}}$.
Figure~\ref{fig:coneprojection} illustrates the notations and the representation of a cone by its corresponding basis and apex.
Thus we associate to each dominance cone $\calL(S_i)$ its corresponding ball basis $B_i=\mathrm{Ball}(\sigma(S_i),r_i)$   on the subspace $H_0$ of zero trace matrices:
$\sigma_i=\sigma(S_i) =  S_i - \frac{\trace(S_i)}{d}I$,
$r_i=r(S_i)=\trace(S_i)\sqrt{1-\frac{1}{d}}$.
We have the following containment relationships:
$P \succ Q \Leftrightarrow \calL(P)\supset  \calL(Q) \Leftrightarrow B(P) \supset B(Q)$
and
$P \succeq Q \Leftrightarrow \calL(P)\supseteq  \calL(Q) \Leftrightarrow B(P) \supseteq B(Q)$
 
Finally, we transform this minimum enclosing {\em matrix} ball problem into a minimum enclosing {\em vector} ball problem using a half-vectorization that preserves the notion of distances, {\it i.e.}, using an isomorphism between the space of symmetric matrices and the space of half-vectorized matrices.
The $\ell_2$-norm of the vectorized matrix should match the matrix Fr\"obenius norm: $\|s\|_2=\|\vec^+(S)\|_2=\|S\|_F$.
Since $\|S\|_F  = \sqrt{ \sum_{i=1}^d  \sum_{j=1}^d  s_{i,j}^2  }= \sqrt{ \sum_{i=1}^d s_{i,i}^2 + 2 \sum_{i=1}^{d-1} \sum_{j=i+1}^d s_{i,j}^2  } =\|   s \|_2$,
it follows that $s = \|\vec^+(S)\|_2= \left[s_{1,1}\ \ldots\ s_{d,d}\ \sqrt{2}s_{1,2}\ \sqrt{2}s_{1,d} \ldots \sqrt{2}s_{d-1,d} \right]^\top \in\bbR^{\frac{d(d+1)}{2}}$.
We can convert back a vector $v\in\bbR^D$ into a corresponding symmetric matrix.

Since we have considered all dominance cones with basis rooted on $H_0^+: \tr(X)\geq 0$ in order to compute the ball basis as orthogonal projections, 
 we need to {\em pre-process} the symmetric matrices to ensure that property as follows:
Let $t=\min\{\tr(S_1), \ldots, \tr(S_n)\}$ denote the minimal trace of the input set of symmetric matrices $S_1, \ldots, S_n$, and define
$S_i'=S_i-tI$ for $i\in [n]$ where $I$ denotes the identity matrix.  Recall that $\tr(X_1+\lambda X_2)=\tr(X_1)+\lambda\tr(X_2)$.
By construction, the transformed input set satisfies $\tr(S_i')\geq 0, \forall i\in [n]$.
Furthermore, observe that $S\succeq S_i$ iff. $S'\succeq S_i'$ where $S'=S-tI$, so that
$\max(S_1, \ldots, S_n) = \max(S_1', \ldots, S_n')+t I$.

As a side note, let us point out that  the reverse basis-sphere-to-cone mapping has been used to compute the convex hull of $d$-dimensional spheres (convex homothets) from
the convex hull of $(d+1)$-dimensional equivalent points~\cite{CHSphere-1996,CHSphere-2003}.

Finally, let us notice that there are severals ways to majorize/minorize matrices:
For example, once can seek  extremal matrices that are invariant up to an {\em invertible transformation}~\cite{QI-switchedsystems-2015}, a stronger requirement than the invariance by orthogonal transformation. In the latter case, it amounts to geometrically compute the Minimum Volume Enclosing Ellipsoid of Ellipsoids (MVEEE)~\cite{QI-switchedsystems-2015,mvee-2008}.

\subsection{Defining $(1+\epsilon)$-approximations of $\bar{S}$}
First, let us summarize the algorithm for computing the L\"owner maximal matrix of a set of $n$ symmetric matrices $S_1, \ldots, S_n$ as follows:
\begin{enumerate}
	\item Normalize matrices so that they have all non-negative traces:
	$$S_i'=S_i-tI, \quad t=\min\{\tr(S_1), \ldots, \tr(S_n)\}.$$
	
	\item Compute the vector ball representations of the dominance cones:
	$$B_i=\mathrm{Ball}\left(\sigma_i,r_i\right)$$ with $$\sigma_i=\vec^+\left(S_i' - \frac{\trace(S_i')}{d}I\right)$$ and $$r_i=\trace(S_i')\sqrt{1-\frac{1}{d}}$$
	
  \item Compute  the small(est) enclosing ball $B'=\mathrm{Ball}(\sigma',r')$ of basis balls (either exactly or an approximation):
	$$
	B'=  \mathrm{Small(est)EnclosingBall}(B_1, \ldots, B_n)
	$$

 \item Convert back the small(est) enclosing ball $B'$ to the dominance cone, and recover its apex $S'$:
$$\bar{S'}=\sigma' + \frac{{r'}}{d}\frac{I}{\sqrt{1-\frac{1}{d}}}.$$
\item Adjust back the matrix trace:
$$\bar{S}= \bar{S'} +  tI, \quad t=\min\{\tr(S_1), \ldots, \tr(S_n)\}.$$

\end{enumerate}

Computing {\em exactly} the extremal L\"owner matrices suffer from the {\em curse of dimensionality} of computing MEBs~\cite{Fisher-2003}.
In~\cite{MatrixData-2007}, Burgeth et al. proceed by discretizing the basis spheres by sampling\footnote{In 2D, we sample $v=[\cos \theta, \sin\theta]^\top$ for $\theta\in [0,2\pi[$. In 3D, we use spherical coordinates $v=[\sin \theta\cos\phi,  \sin\theta\sin\phi,  \cos\theta]^\top$ for $\theta\in [0,2\pi[$ and $\phi\in [0,\pi[$. } the extreme  x points $v v^\top$ for $\|v\|_2=1$.
This yields an approximation term, requires more computation, and even worse the method  does not scale~\cite{SEBB-2004} in high-dimensions.
Thus in order to handle high-dimensional matrices met in software formal verification~\cite{QI-switchedsystems-2015} or in computer vision (structure tensor~\cite{Forsner-1986}), 
we consider  $(1+\epsilon)$-approximation of the extremal L\"owner matrices.
The notion of tightness of approximation of $\bar{S}$ (the epsilon) is imported straightforwardly from the definition of the tightness of the geometric covering problems.
A $(1+\epsilon)$-approximation $\tilde{{S}}$ of $\bar S$ is a matrix $\tilde{{S}}\succ \bar S$ such that: $r(\tilde{{S}})\leq (1+\epsilon)r(\bar S)$.
It follows  from Eq.~\ref{eq:matrad} that a $(1+\epsilon)$-approximation satisfies $\tr(\tilde{{S}})\leq (1+\epsilon) \tr(\bar S)$.

We present a fast guaranteed approximation algorithm for approximating the minimum enclosing ball of a set of balls (or more generally, for sets of compact geometric objects).

\section{Approximating the minimum enclosing ball of objects and balls \label{sec:mebb}}

\def\ttt{0.40\columnwidth}
\begin{figure}
\centering
\begin{tabular}{ll}
\includegraphics[width=\ttt]{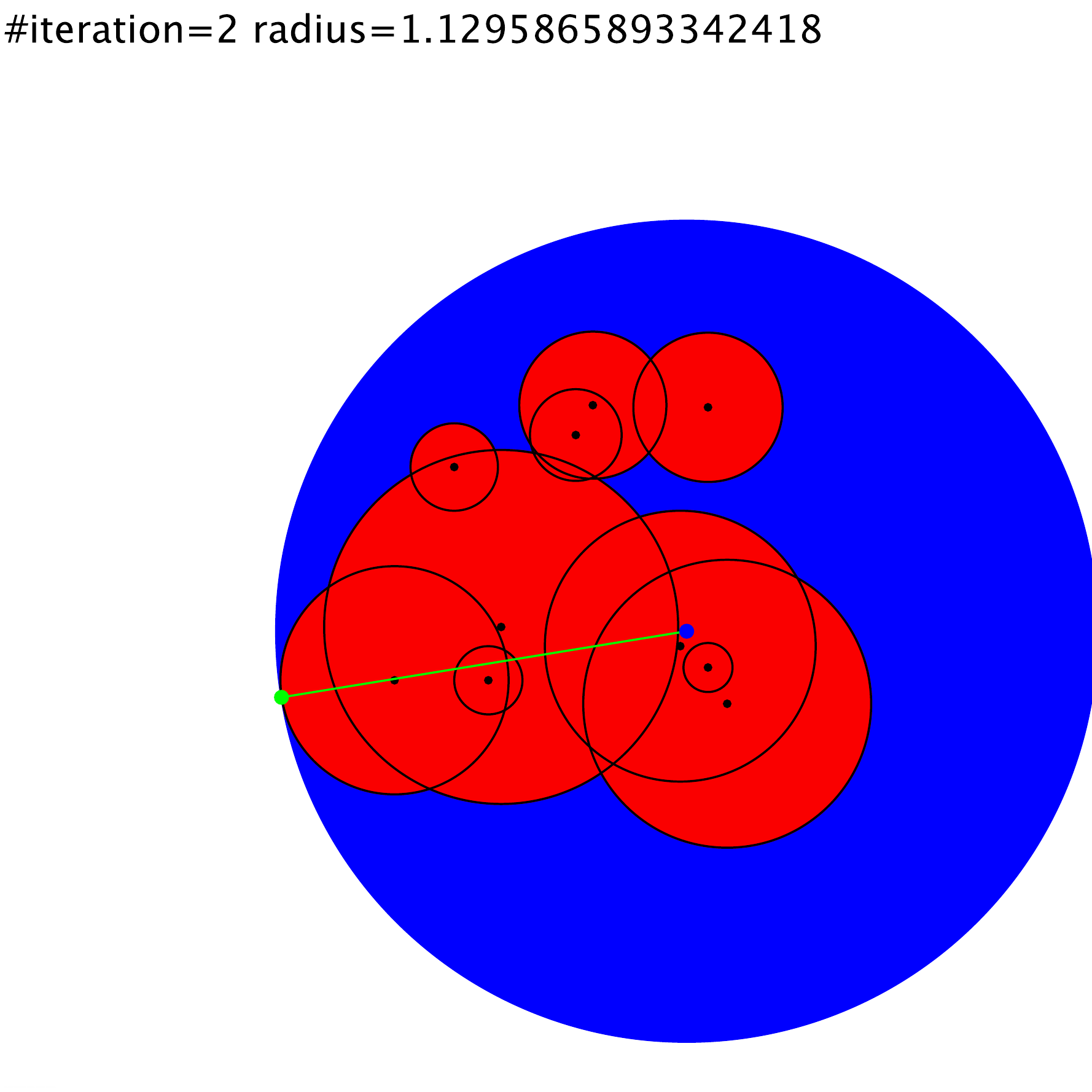}  & \includegraphics[width=\ttt]{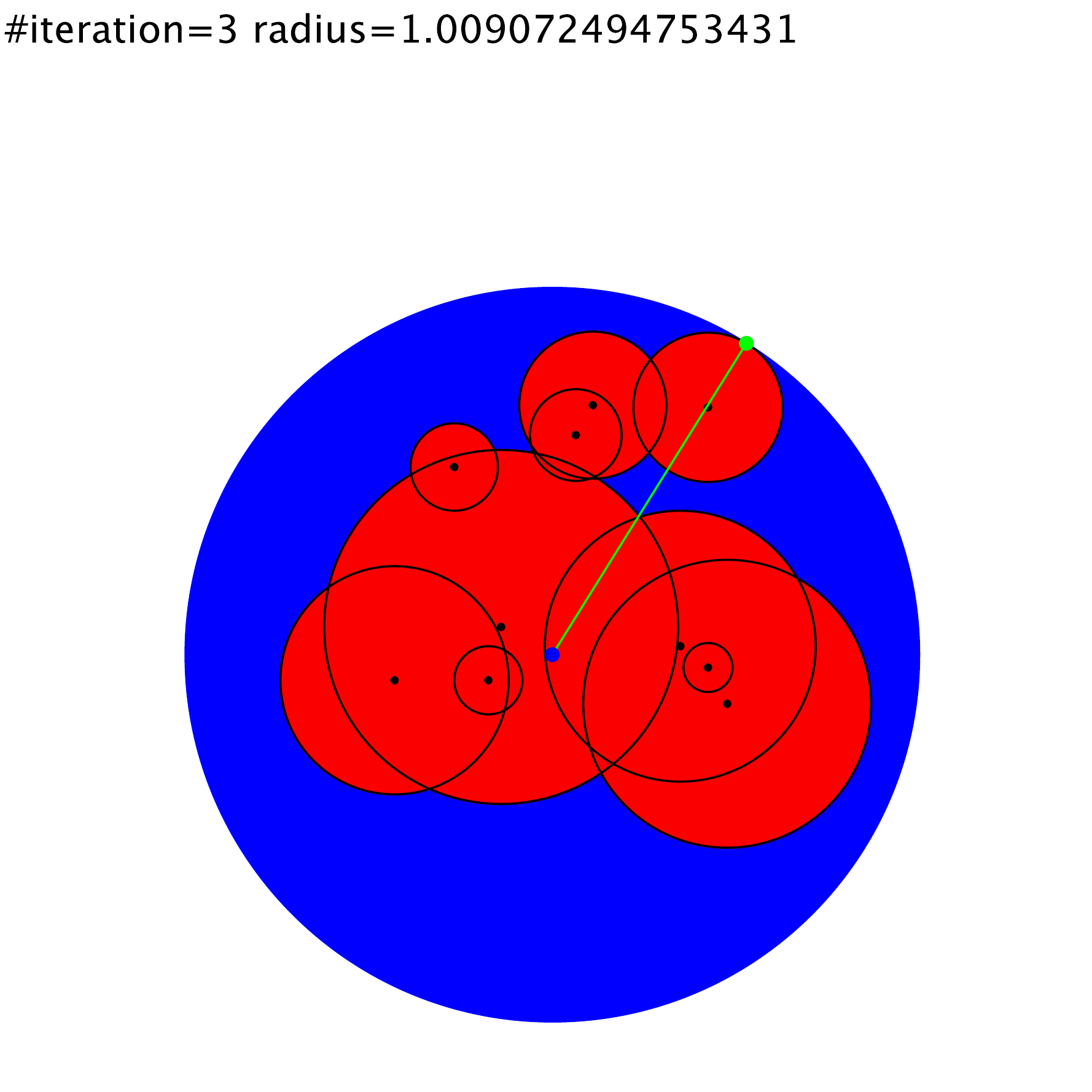}\\
\includegraphics[width=\ttt]{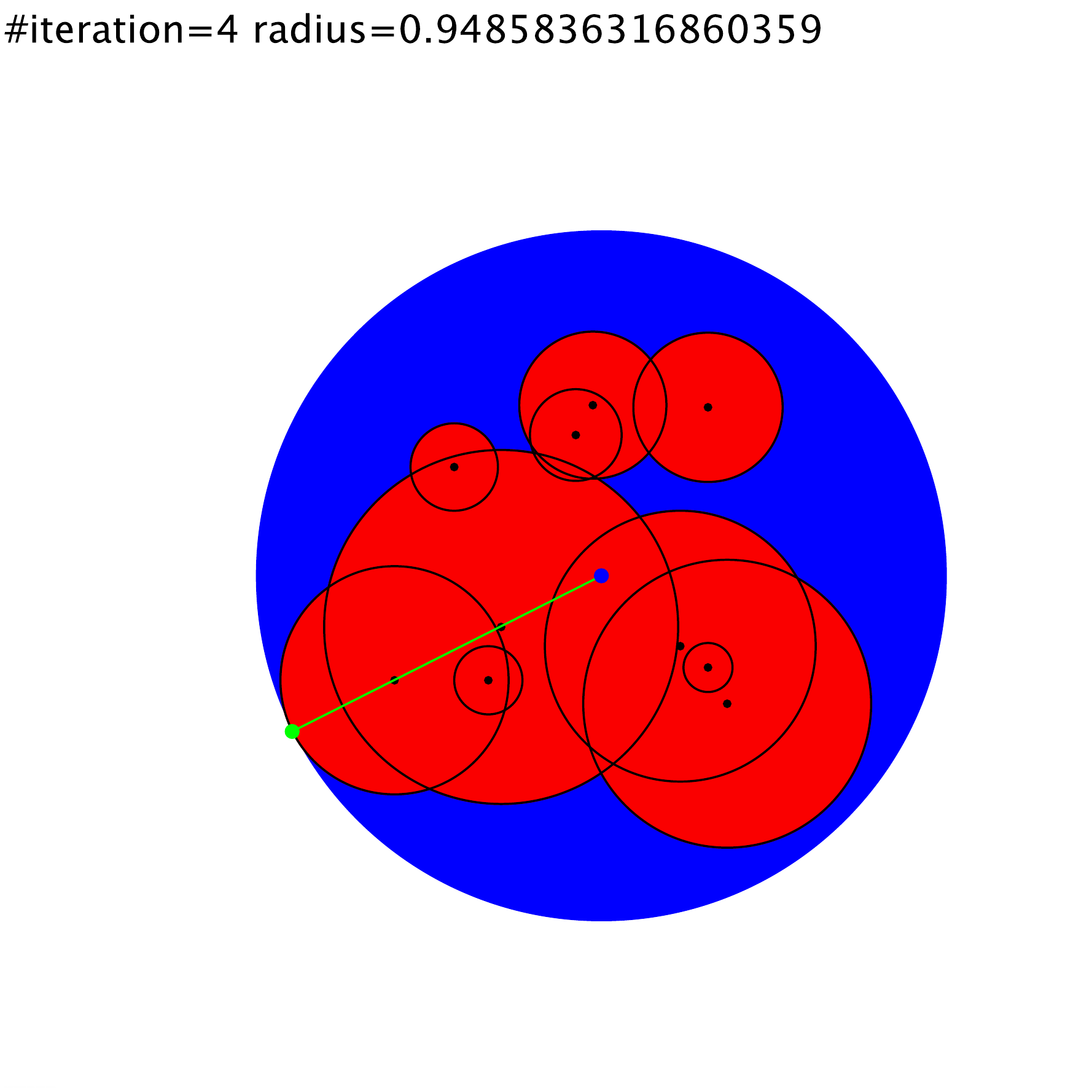}  & \includegraphics[width=\ttt]{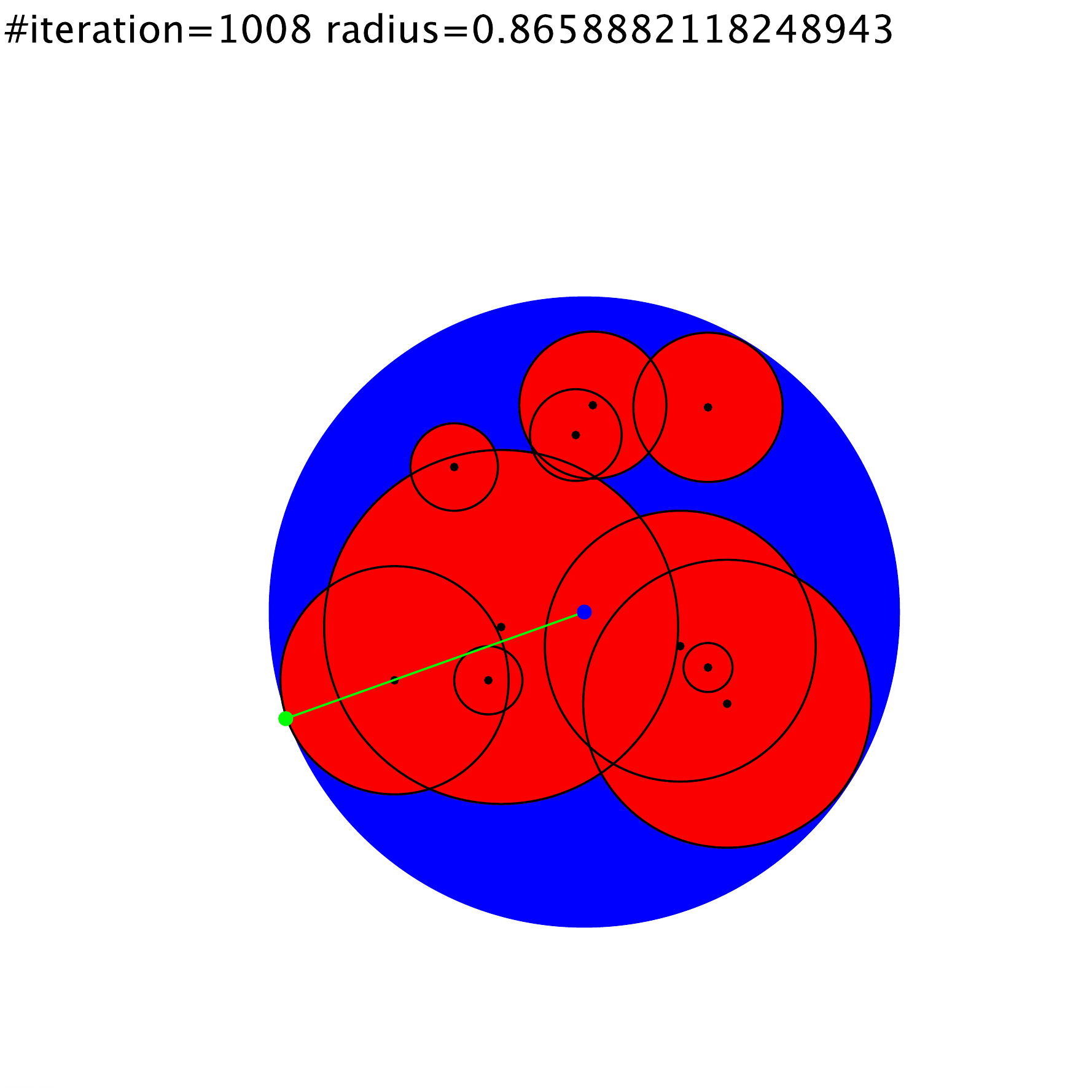}\\
\includegraphics[width=\ttt]{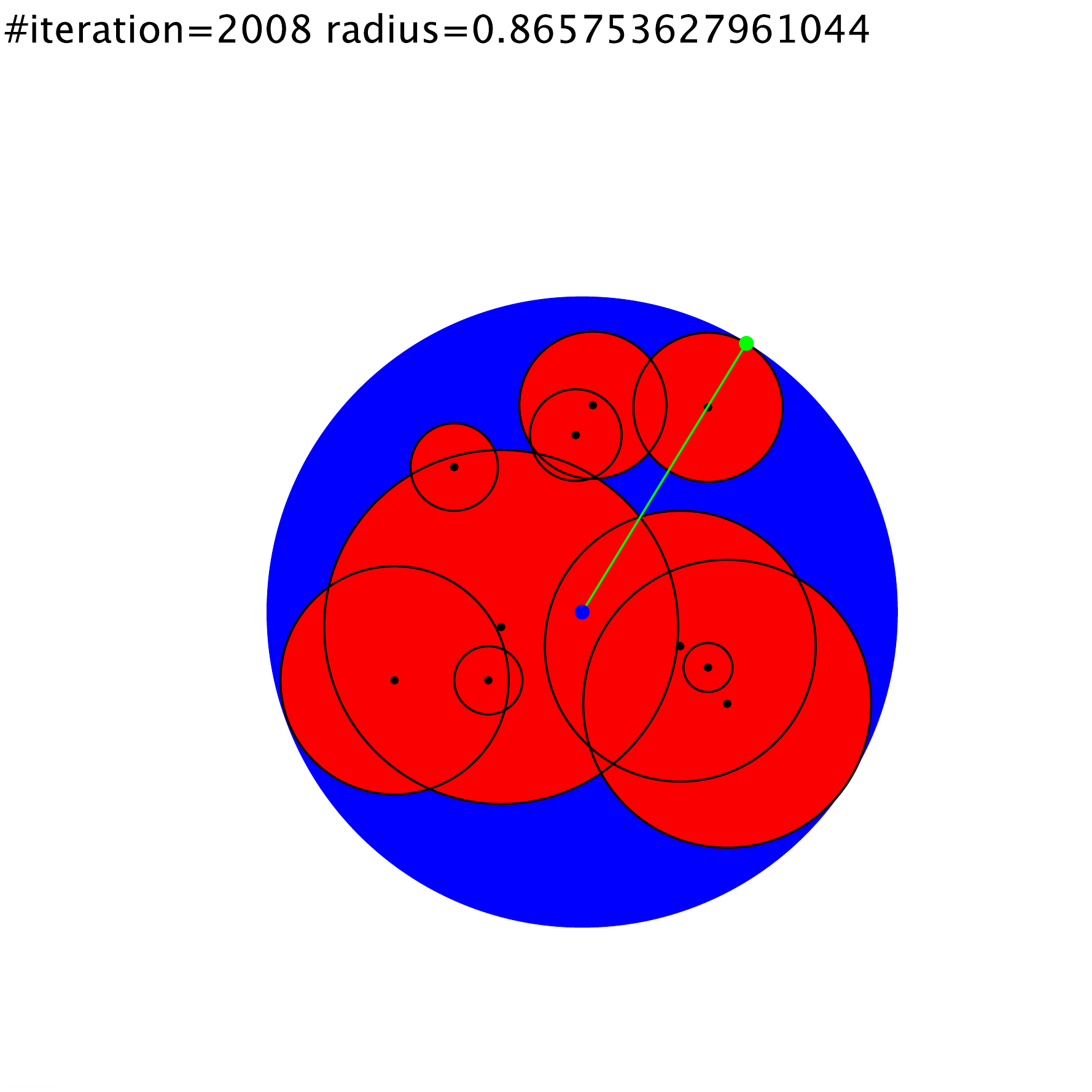}  & \includegraphics[width=\ttt]{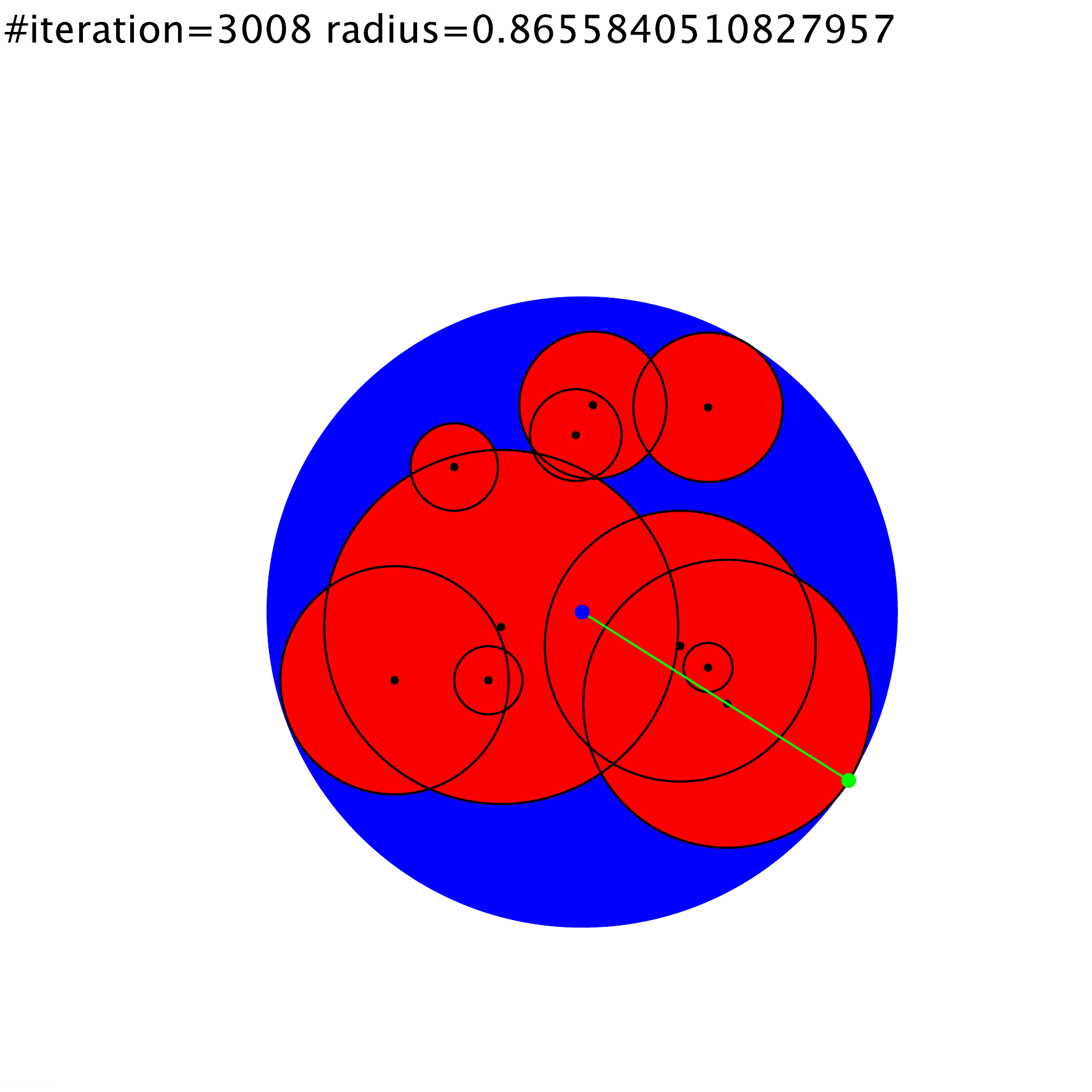}\\
\end{tabular}

\caption{Approximating the minimum enclosing ball of balls iteratively: Snapshots at iterations 1, 2, 3,1008, 2008 and 3008 (best viewed in color).\label{fig:bcb}}
\end{figure}

We extend the incremental algorithm of B\u{a}doiu  and Clarkson~\cite{BC-2008} (BC) designed for finite point sets to  {\it ball sets} or {\it compact  object sets} that work in large dimensions.
Let $B_1=\mathrm{Ball}(c_1,r_1), ..., B_n=\mathrm{Ball}(c_n,r_n)$ denote a set of $n$ balls.
For an object $\calO$ and a query point $q$, denote by $D^f(q,\calO)$ the {\em farthest} distance from $q$ to $\calO$: $D^f(q,\calO)=\max_{o\in \calO} \|q-o\|$, 
and let $F(q,\calO)$ denote the farthest point of $\calO$ from $q$.
The generalized BC ~\cite{BC-2008} algorithm for approximating the circumcenter of the minimum volume enclosing ball of $n$ objects (MVBO) $\calO_1, \ldots, \calO_n$ is summarized as follows:
\begin{itemize}
\item Let $e_1\leftarrow x\in \calO_1$ and $i\leftarrow 1$.
\item Repeat $l$ times: 
\begin{itemize}
 \item Find the farthest object $\calO_f$  to current center: $f=\arg\max_{j\in [n]} D^f(e_i,\calO_j)$ 
 \item Update the  circumcenter: $e_{i+1} = \frac{i}{i+1}e_i + \frac{1}{i+1} (F(e_i,\calO_f)-e_i)$
\item $i\leftarrow i+1$.
\end{itemize}
\end{itemize}
When considering balls as objects, the farthest distance of a point $x$ to a ball $B_j=\ball(c_j,r_j)$ is $D^f(e_i,B_j) = \| c_j-e_i\| + r_j$, and the circumcenter updating rule is:
$e_{i+1} = \frac{i}{i+1}e_i + \frac{1}{i+1} (c_f-e_i) \left(1+\frac{r_f}{\| c_f-e_i\|}\right)$.
See Figure~\ref{fig:bcb} and online video\footnote{\protect\url{https://www.youtube.com/watch?v=w1ULgGAK6vc}} for an illustration.
(MVBO can also be used to approximate the MEB of ellipsoids.)
It is proved in~\cite{BC-2003} that at iteration $i$, we have $\|e_i-e^*\|\leq \frac{r^*}{\sqrt{i}}$ where $B^*=\mathrm{Ball}(e^*,r^*)$ is the unique smallest enclosing ball. Hence the radius of the ball centered at $e_i$ is bounded by $(1+\frac{1}{\sqrt{i}})r^*$. To get a $(1+\epsilon)$-approximation, we need $\frac{1}{\epsilon^2}$ iterations.s
It follows that a $(1+\epsilon)$-approximation of the smallest enclosing ball of $n$ $D$-dimensional balls can be computed in $O(\frac{D}{n}{\epsilon^2})$-time~\cite{BC-2003}, and since $D=O(d^2)$ we get:
\begin{theorem}
The L\"owner maximal matrix $\bar{S}$ of a set of $n$  $d$-dimensional symmetric matrices can be approximated by a matrix $\tilde{S}\succ\bar{S}$ such that
 $\tr(\tilde{{S}})\leq (1+\epsilon) \tr(\bar S)$ in $O(\frac{d^2}{n}{\epsilon^2})$-time. 
\end{theorem}
Interestingly, this shows that the approximation of L\"owner supremum matrices admits core-sets~\cite{BC-2003}, the subset of farthest balls $B_{f(i)}$ chosen during the $l$ iterations, so that $\tilde{S}=\max(S_{f(1)}, \ldots, S_{f(l)})$ with $\tr(\tilde{{S}})\leq (1+\epsilon) \tr(\bar S)$.
See~\cite{ameb-2003} for other MEB approximation algorithms.

To a symmetric matrix $S$, we associate a {\em quadratic form} $q_S(x)=x^\top S x$ that is a strictly convex function when $S$ is PSD.   
Therefore, we  may visualize the SPSD matrices in 2D/3D as ellipsoids (potentially degenerated flat ellipsoids for rank-deficient matrices). 
More precisely, we associate to each positive  definite matrix $S$, a geometric ellipsoid  defined by 
$\calE(S) = \{ x\in\bbR^d \st  x^\top S^{-1} x = \rho\}$,
where $\rho$ is a prescribed constant (usually set to $\rho=1$, Figure~\ref{fig:viz}).
From the SVD decomposition of $S^{-1}$, we recover the rotation matrix, and the semi-radii of the ellipsoid are the 
square root eigenvalues $\sqrt{\lambda_1}, \ldots, \sqrt{\lambda_d}$.
It follows that $P\succeq Q  \Leftrightarrow  \calE(P) \supseteq  \calE(Q)$.
To handle degenerate flat ellipsoids that are not fully dimensional (rank-deficient matrix $P$), we define $\calE(P) = \{ x\in\bbR^d \st x x\top \preceq P \}$.
Note that those ellipsoids are all centered at the origin, and may also conceptually be thought as centered Gaussian distributions (or covariance matrices denoting
the concentration ellipsoids of estimators~\cite{AppLoewner-1967} in statistics).
We  can also visualize the L\"owner ordering cone and dominance cones for $2\times 2$ matrices embedded in the vectorized 3D space of symmetric matrices (Figure~\ref{fig:viz}), and the corresponding  
half-vectorized ball basis (Figure~\ref{fig:viz}).

\def\ttt{0.32\columnwidth}
\begin{figure}
\centering
\begin{tabular}{ccc}
\includegraphics[width=\ttt]{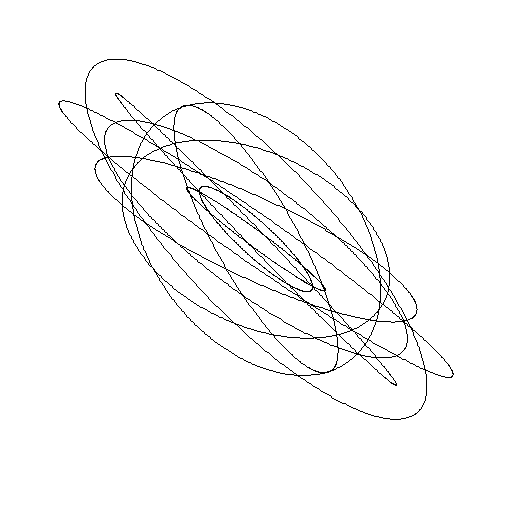} &
\includegraphics[width=\ttt]{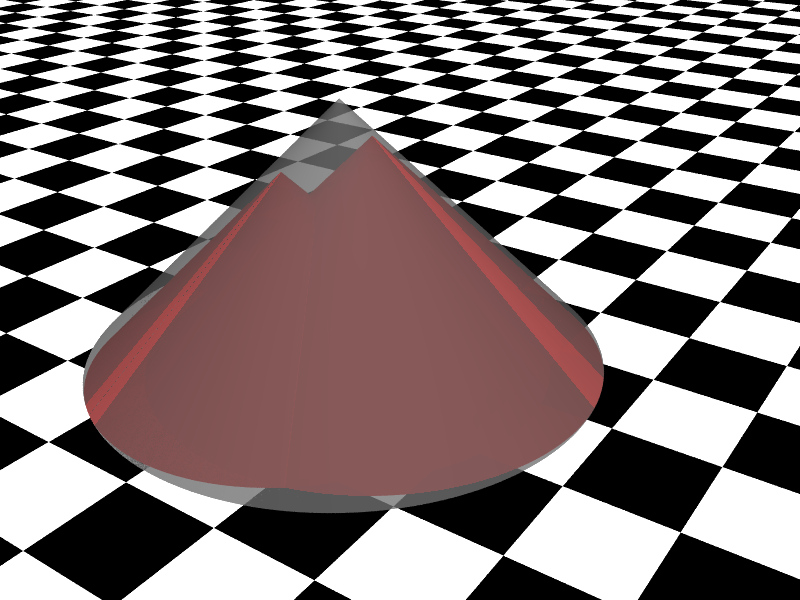} &
\includegraphics[width=\ttt]{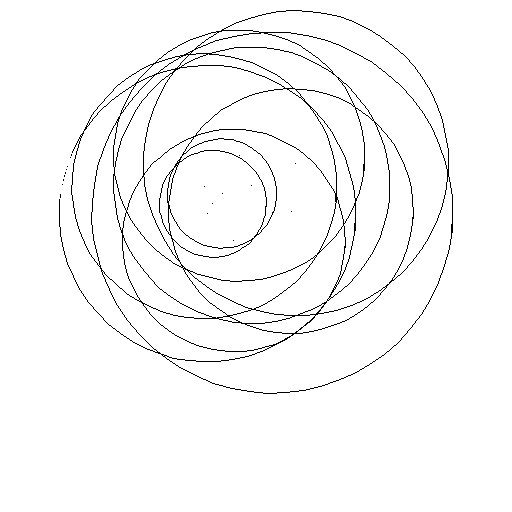} \\
(a) & (b) & (c)
\end{tabular}

\caption{Equivalent visualizations: (a) $2\times 2$ PSD matrices visualized as ellipsoids, with (b) corresponding 3D vector L\" owner cones, and (c) corresponding cone vector ball basis.
\label{fig:viz}}
\end{figure}

\section{Concluding remarks\label{sec:concl}}

Our novel extremal matrix approximation method allows one to leverage  further related results related to  core-sets~\cite{BC-2008} for dealing with high-dimensional extremal matrices.
For example, we may consider clustering PSD matrices with respect to L\"owner order and use the $k$-center clustering technique with guaranteed approximation~\cite{kcenter-2003,coreset-clustering-2009}. 
A Java\texttrademark{} code of our method is available for reproducible research.

\section*{Acknowledgements}
This work was carried out during the Matrix Information Geometry (MIG) workshop~\cite{mig-2013}, organized at \'Ecole Polytechnique, France in February 2011 (\url{https://www.sonycsl.co.jp/person/nielsen/infogeo/MIG/}).
Frank Nielsen dedicates this work to the memory of his late father Gudmund Liebach Nielsen who passed away during the last day of the workshop.


\begin{thebibliography}{10}
\providecommand{\url}[1]{#1}
\csname url@samestyle\endcsname
\providecommand{\newblock}{\relax}
\providecommand{\bibinfo}[2]{#2}
\providecommand{\BIBentrySTDinterwordspacing}{\spaceskip=0pt\relax}
\providecommand{\BIBentryALTinterwordstretchfactor}{4}
\providecommand{\BIBentryALTinterwordspacing}{\spaceskip=\fontdimen2\font plus
\BIBentryALTinterwordstretchfactor\fontdimen3\font minus
  \fontdimen4\font\relax}
\providecommand{\BIBforeignlanguage}[2]{{%
\expandafter\ifx\csname l@#1\endcsname\relax
\typeout{** WARNING: IEEEtran.bst: No hyphenation pattern has been}%
\typeout{** loaded for the language `#1'. Using the pattern for}%
\typeout{** the default language instead.}%
\else
\language=\csname l@#1\endcsname
\fi
#2}}
\providecommand{\BIBdecl}{\relax}
\BIBdecl

\bibitem{Bathia-2009}
R.~Bhatia, \emph{Positive definite matrices}.\hskip 1em plus 0.5em minus
  0.4em\relax Princeton university press, 2009.

\bibitem{AppLoewner-1967}
M.~Siotani, ``Some applications of {L}oewner's ordering on symmetric
  matrices,'' \emph{Annals of the Institute of Statistical Mathematics},
  vol.~19, no.~1, pp. 245--259, 1967.

\bibitem{Angulo-2013}
J.~Angulo, ``Supremum/infimum and nonlinear averaging of positive definite
  symmetric matrices,'' \emph{Matrix Information Geometry}, pp. 3--33, 2013.

\bibitem{PSDMorpho-2007}
B.~Burgeth, A.~Bruhn, N.~Papenberg, M.~Welk, and J.~Weickert, ``Mathematical
  morphology for matrix fields induced by the {L}oewner ordering in higher
  dimensions,'' \emph{Signal Processing}, vol.~87, 2007.

\bibitem{QI-switchedsystems-2015}
X.~Allamigeon, S.~Gaubert, E.~Goubault, S.~Putot, and N.~Stott, ``A scalable
  algebraic method to infer quadratic invariants of switched systems,'' in
  \emph{Embedded Software (EMSOFT), 2015 International Conference on}, Oct
  2015, pp. 75--84.

\bibitem{LownerMLE-1991}
J.~A. Calvin and R.~L. Dykstra, ``Maximum likelihood estimation of a set of
  covariance matrices under {L}\"owner order restrictions with applications to
  balanced multivariate variance components models,'' \emph{The Annals of
  Statistics}, pp. 850--869, 1991.

\bibitem{MLE-Lowner-2007}
M.-T. Tsai, ``Maximum likelihood estimation of {W}ishart mean matrices under
  {L}{\"o}wner order restrictions,'' \emph{Journal of Multivariate Analysis},
  vol.~98, no.~5, pp. 932--944, 2007.

\bibitem{Forsner-1986}
W.~F\"{o}rstner, ``{A Feature Based Correspondence Algorithm for Image
  Matching},'' \emph{Int. Arch. of Photogrammetry and Remote Sensing}, vol.~26,
  no.~3, pp. 150--166, 1986.

\bibitem{MatrixData-2007}
B.~Burgeth, A.~Bruhn, S.~Didas, J.~Weickert, and M.~Welk, ``Morphology for
  matrix data: Ordering versus {PDE}-based approach,'' \emph{Image and Vision
  Computing}, vol.~25, no.~4, pp. 496--511, 2007.

\bibitem{SPSDCone-1987}
R.~D. Hill and S.~R. Waters, ``On the cone of positive semidefinite matrices,''
  \emph{Linear Algebra and its Applications}, vol.~90, pp. 81--88, 1987.

\bibitem{CHSphere-1996}
J.-D. Boissonnat, A.~C{\'e}r{\'e}zo, O.~Devillers, J.~Duquesne, and M.~Yvinec,
  ``An algorithm for constructing the convex hull of a set of spheres in
  dimension $d$,'' \emph{Computational Geometry}, vol.~6, no.~2, pp. 123--130,
  1996.

\bibitem{CHSphere-2003}
J.-D. Boissonnat and M.~I. Karavelas, ``On the combinatorial complexity of
  euclidean {V}oronoi cells and convex hulls of $d$-dimensional spheres,'' in
  \emph{Proceedings of the fourteenth annual ACM-SIAM symposium on Discrete
  algorithms}.\hskip 1em plus 0.5em minus 0.4em\relax Society for Industrial
  and Applied Mathematics, 2003, pp. 305--312.

\bibitem{mvee-2008}
S.~Jambawalikar and P.~Kumar, ``A note on approximate minimum volume enclosing
  ellipsoid of ellipsoids,'' in \emph{Computational Sciences and Its
  Applications, 2008. ICCSA'08. International Conference on}.\hskip 1em plus
  0.5em minus 0.4em\relax IEEE, 2008, pp. 478--487.

\bibitem{Fisher-2003}
K.~Fischer, B.~G{\"a}rtner, and M.~Kutz, ``Fast smallest-enclosing-ball
  computation in high dimensions,'' in \emph{Algorithms-ESA 2003}.\hskip 1em
  plus 0.5em minus 0.4em\relax Springer, 2003, pp. 630--641.

\bibitem{SEBB-2004}
K.~Fischer and B.~G{\"a}rtner, ``The smallest enclosing ball of balls:
  combinatorial structure and algorithms,'' \emph{International Journal of
  Computational Geometry \& Applications}, vol.~14, no. 04n05, pp. 341--378,
  2004.

\bibitem{BC-2008}
M.~B{\u{a}}doiu and K.~L. Clarkson, ``Optimal core-sets for balls,''
  \emph{Computational Geometry}, vol.~40, no.~1, pp. 14--22, 2008.

\bibitem{BC-2003}
\BIBentryALTinterwordspacing
------, ``Smaller core-sets for balls,'' in \emph{Proceedings of the Fourteenth
  Annual ACM-SIAM Symposium on Discrete Algorithms}, ser. SODA '03.\hskip 1em
  plus 0.5em minus 0.4em\relax Philadelphia, PA, USA: Society for Industrial
  and Applied Mathematics, 2003, pp. 801--802. [Online]. Available:
  \url{http://dl.acm.org/citation.cfm?id=644108.644240}
\BIBentrySTDinterwordspacing

\bibitem{ameb-2003}
P.~Kumar, J.~S. Mitchell, and E.~A. Yildirim, ``Approximate minimum enclosing
  balls in high dimensions using core-sets,'' \emph{Journal of Experimental
  Algorithmics (JEA)}, vol.~8, pp. 1--1, 2003.

\bibitem{kcenter-2003}
J.~Mihelic and B.~Robic, ``Approximation algorithms for the $k$-center problem:
  An experimental evaluation,'' in \emph{Selected papers of the International
  Conference on Operations Research (SOR 2002)}.\hskip 1em plus 0.5em minus
  0.4em\relax Springer, 2003, p. 371.

\bibitem{coreset-clustering-2009}
K.~Chen, ``On coresets for $k$-median and $k$-means clustering in metric and
  euclidean spaces and their applications,'' \emph{SIAM Journal on Computing},
  vol.~39, no.~3, pp. 923--947, 2009.
	
	\bibitem{mig-2013}
F. Nielsen and R. Bathia, 
``Matrix Information Geometry,'' Springer, 2013.
\url{http://www.springer.com/fr/book/9783642302312}	

\end{thebibliography}

\end{document}